\documentclass[review]{elsarticle}

\usepackage{lineno,hyperref}
\usepackage{caption}
\usepackage{subcaption}
\modulolinenumbers[5]

\usepackage[utf8]{inputenc}
\usepackage[T1]{fontenc}
\usepackage{graphicx}
\usepackage{enumitem}
\usepackage{amssymb}

\journal{Plasma Physics and Controlled Fusion}

\begin{document}

\begin{frontmatter}

\title{Analysis of the neutral fluxes in the divertor region of Wendelstein 7-X under attached and detached conditions using EMC3-EIRENE}


\author[affiliation1]{Dieter Boeyaert \corref{mycorrespondingauthor}}
\cortext[mycorrespondingauthor]{Corresponding author}
\ead{boeyaert@wisc.edu}

\author[affiliation2]{Yuhe Feng}

\author[affiliation1]{Heinke Frerichs}

\author[affiliation2]{Thierry Kremeyer}

\author[affiliation2]{Dirk Naujoks}

\author[affiliation2]{Felix Reimold}

\author[affiliation1]{Oliver Schmitz}

\author[affiliation2]{Victoria Winters}

\author[affiliation2]{Sergey Bozhenkov}

\author[affiliation2]{Joris Fellinger}

\author[affiliation2]{Marcin Jakubowski}

\author[affiliation2]{Ralf König}

\author[affiliation2]{Maciej Krychowiak}

\author[affiliation2]{Valeria Perseo}

\author[affiliation2]{Georg Schlisio}

\author[affiliation2]{Uwe Wenzel}

\author[ref3]{W7-X team}

\address[affiliation1]{University of Wisconsin-Madison, Department of Engineering Physics, 1500 Engineering Drive, Madison, WI 53706, United States of America}
\address[affiliation2]{Max-Planck-Institut f\"{u}r Plasmaphysik, Teilinstitut Greifswald, Wendelsteinstrasse 1, 17491 Greifswald, Germany}
\address[ref3]{The  full  list  of  W7-X  team  members  is  given  in  Sunn  Pedersen, T.  et  al., Nucl. Fusion 62 (2022) 042022}

\begin{abstract}
	
	This paper analyzes the neutral fluxes in the divertor region of the W7-X standard configuration for different input powers, both under attached and detached conditions. The performed analysis is conducted through EMC3-EIRENE simulations. They show the importance of the horizontal divertor to generate neutrals, and resolve the neutral plugging in the divertor region. Simulations of detached cases show a decrease in the number of generated neutrals compared to the attached simulations, in addition to a higher fraction of the ion flux arriving on the baffles during detachment. As the ionization takes place further inside the plasma during detachment, a larger percentage of the generated neutral particles leave the divertor as neutrals. The leakage in the poloidal and toroidal direction increases, just as the fraction of collected particles at the pumping gap. The fraction of pumped particles increases with a factor two, but stays below one percent. This demonstrates that detachment with the current target geometry, although it improves the power exhaust, is not yet leading to an increased particle exhaust. 

\end{abstract}

\begin{keyword}
EMC3-EIRENE\sep W7-X\sep neutrals \sep divertor \sep plugging
\end{keyword}

\end{frontmatter}


\section{Introduction}

For the operation of fusion power plants based on magnetic confinement, both tokamak and stellarator concepts are currently studied. An advantage of a stellarator in comparison with a tokamak is the inherent capability for steady-state operation due to the missing requirement of an induced plasma current. In all fusion power plants, density control and exhaust of neutral particles are crucial for operation \cite{day2014development}. These neutrals can originate from the helium ash due to neutralization on first wall surfaces and targets, or through recombination, or as extrinsic fueled particles \cite{reiter1990burn}. As no fusion is taking place in current devices, and as in W7-X no recombination was observed \cite{schmitz2020stable}, the present neutrals are only originating from plasma-wall interactions at the plasma facing components, and extrinsic fueling.

\begin{figure}[h]
	\centering
		\begin{subfigure}{12cm}
		\includegraphics[width=12cm]{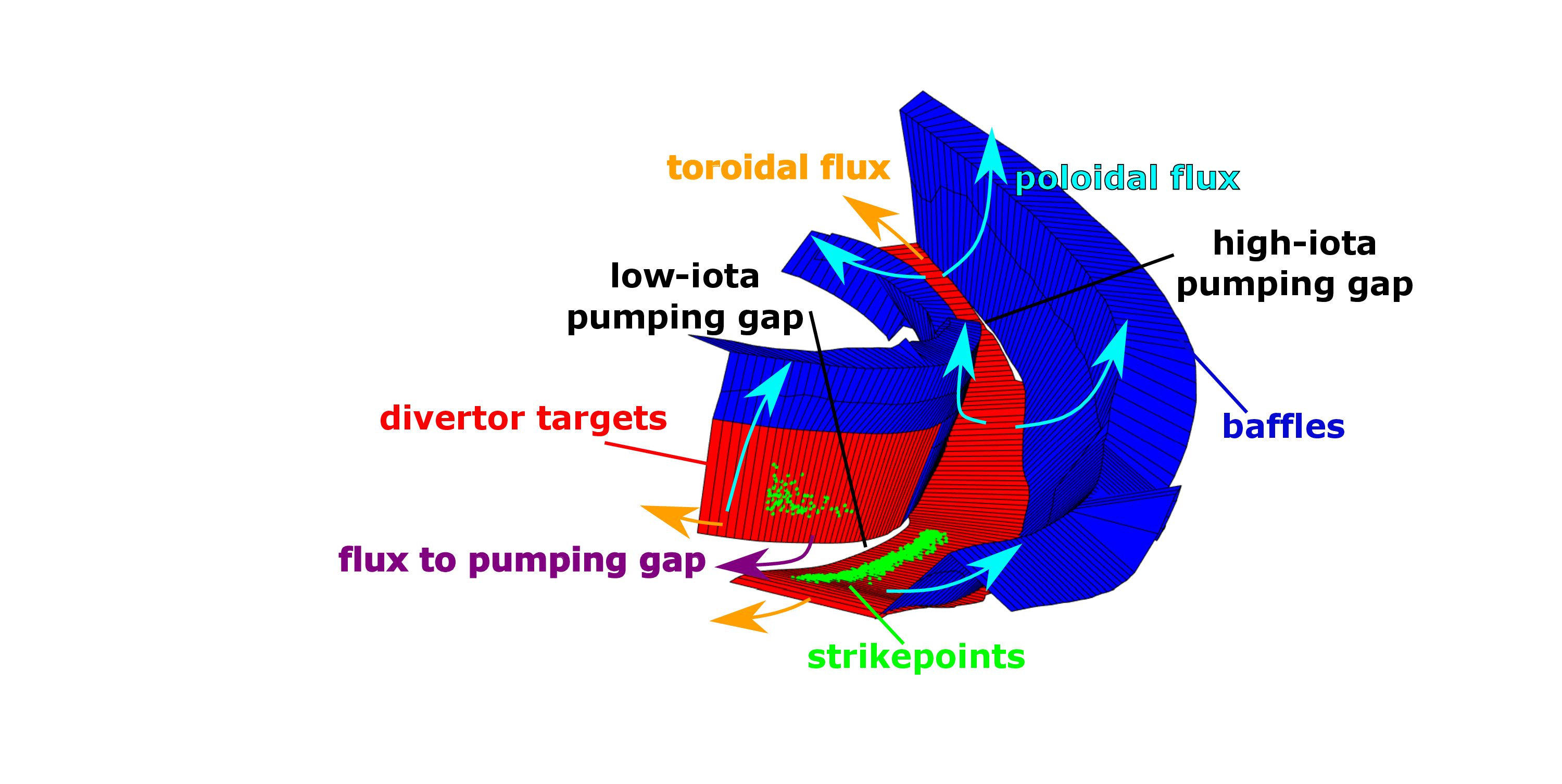}
		\caption{}
		\label{subfig:divertor_region_overview}
	\end{subfigure}
	\begin{subfigure}{6cm}
		\includegraphics[width=6cm]{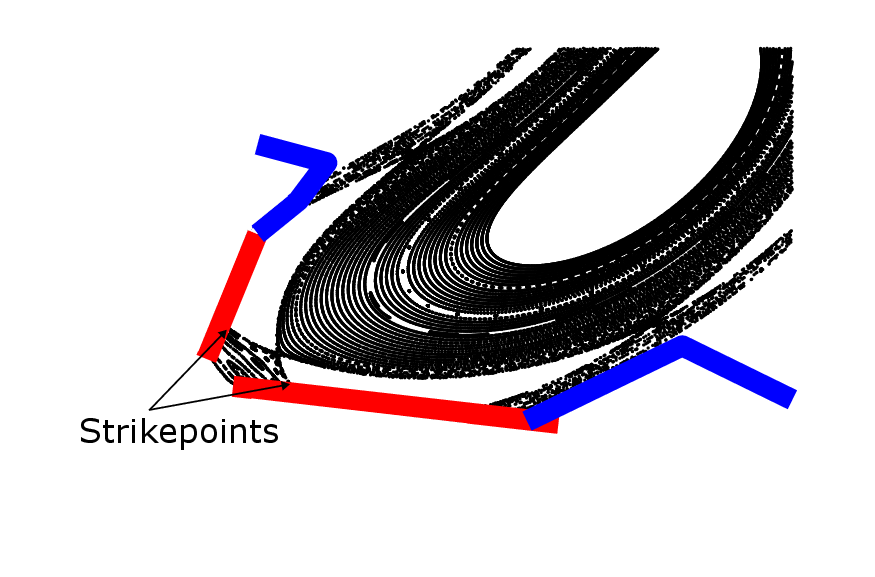}
		\caption{}
		\label{subfig:divertor_detail}
	\end{subfigure}
	\caption{The different paths a neutral can follow if they are not ionized inside the divertor region. In the presented work, the poloidal and toroidal neutral fluxes are evaluated as shown in the figure, as well as the flux through the pumping gap. In (a) one divertor module is shown. In (b) a toroidal cross section in the low-iota region (region where the low-iota pumping gap is located) is shown together with the Poincar{\'e} plot at this location. In both figures the baffles are indicated in blue and the divertor targets in red.}
	\label{fig:divertor_region}
\end{figure}

Wendelstein 7-X (W7-X) is a stellarator which is optimized for reduced neoclassical transport \cite{pedersen2022experimental}. W7-X exists of five identical modules from which each module is in itself stellarator symmetric. To enhance power and particle exhaust from the core plasma, a divertor is installed. The employed divertor is an island-divertor meaning that the divertor plates are intersecting large magnetic islands in the edge \cite{strumberger1996sol}. The plasma facing components in W7-X are made from carbon which causes carbon impurities in the plasma. In figure \ref{subfig:divertor_region_overview} the divertor region of one W7-X module is shown. In this figure the divertor targets are indicated in red and the surrounding baffles in blue. The resulting strikepoints for the investigated configuration are calculated with field line tracing using FLARE \cite{frerichs2015field} and shown in green. The effect of the employed island configuration on the Poincaré map in the low-iota region can be seen in figure \ref{subfig:divertor_detail}. 

W7-X can operate in different magnetic configurations. The configuration which is investigated in this paper and from which the Poincaré map and strikepoints are shown in figure \ref{fig:divertor_region}, is the so-called standard configuration in which the same current is used in all non-planar coils and no current is applied to the planar coils \cite{geiger2014physics}. The effect of other magnetic configurations on the strikepoint locations and the overall plasma performance can be found in ref. \cite{pedersen2018first}. 

In the presented work, the origin and exhaust of neutrals in hydrogen experiments performed during Operation Phase 1.2b (OP1.2b) are analyzed. The origin of the neutrals is determined by the divertor targets as they are the main plasma facing components \cite{kremeyer2022analysis, winters2021emc3}. Refs. \cite{bader2016modeling,stephey2018impact} indicate that the location from the magnetic islands in the plasma can effect hydrogen and helium fueling and exhaust in W7-X, LHD, and HSX. Ref. \cite{stephey2018impact} shows that the magnetic configurations for W7-X examined in that paper do not influence the hydrogen exhaust, but they do influence the helium exhaust. This is important as one of the key goals from a divertor is to remove the helium ash resulting from the fusion reaction. In the current work, however, these effects are not investigated as the focus is on an in-depth analysis of the hydrogen neutrals in the standard configuration of W7-X. Refs. \cite{masuzaki2010design, morisaki2013initial} show for LHD that the position of the divertor targets and baffles determines the subdivertor pressure and in that way the particle exhaust. Modifying the location and shape of the divertor and baffles of W7-X, however, would not allow its current magnetic flexibility. This leads to the baffle design shown in figure \ref{fig:divertor_region} which is called an open divertor structure \cite{renner2002divertor}. The incoming plasma neutralizes at the strikepoint locations. The role of the baffles is to keep the generated neutrals in the divertor region and guide them to the subdivertor. Due the open divertor structure, the neutrals are not trapped as well as would be the case in a closed divertor configuration.


The examined experiments have an input power of 3.3 MW and 5.5 MW. They are selected from the experiments presented in ref. \cite{schmitz2020stable}. Neutrals which are not ionized in the divertor, can leave the divertor region through the pumping gap, in the poloidal direction, or in the toroidal direction as indicated in figure \ref{fig:divertor_region}. A first step to improve the neutral exhaust, is a better understanding of the fluxes indicated in figure \ref{fig:divertor_region}. As these neutral fluxes cannot be measured directly in experiments, EMC3-EIRENE modeling  is used to get a better understanding of their spatial distribution. 

EMC3-EIRENE is a coupling between the EMC3 and the EIRENE code. EMC3 is a kinetic Monte Carlo code solving the Braginskii equations for plasma particles, energy and momentum \cite{feng2014recent}, where EIRENE is a Monte Carlo code solving the Boltzmann equation providing the neutral source terms in the Braginskii equations \cite{reiter2005eirene}.

In a first section the setup of the EMC3-EIRENE simulations is described. Section \ref{sec:comparison_exp_sim} handles the comparison between performed simulations and experimental neutral subdivertor pressures to ensure the simulation mimics well the neutral behavior of the experiments. This allows to investigate the neutral behavior in sections \ref{sec:origin_neutrals} and \ref{sec:exhaust} dealing with the origin and exhaust of the present neutrals, before coming to a summary of the main results in the last section.

\section{Simulation setup}
\label{sec:setup}

\begin{figure}[h]
	\centering
	\includegraphics[width=13cm]{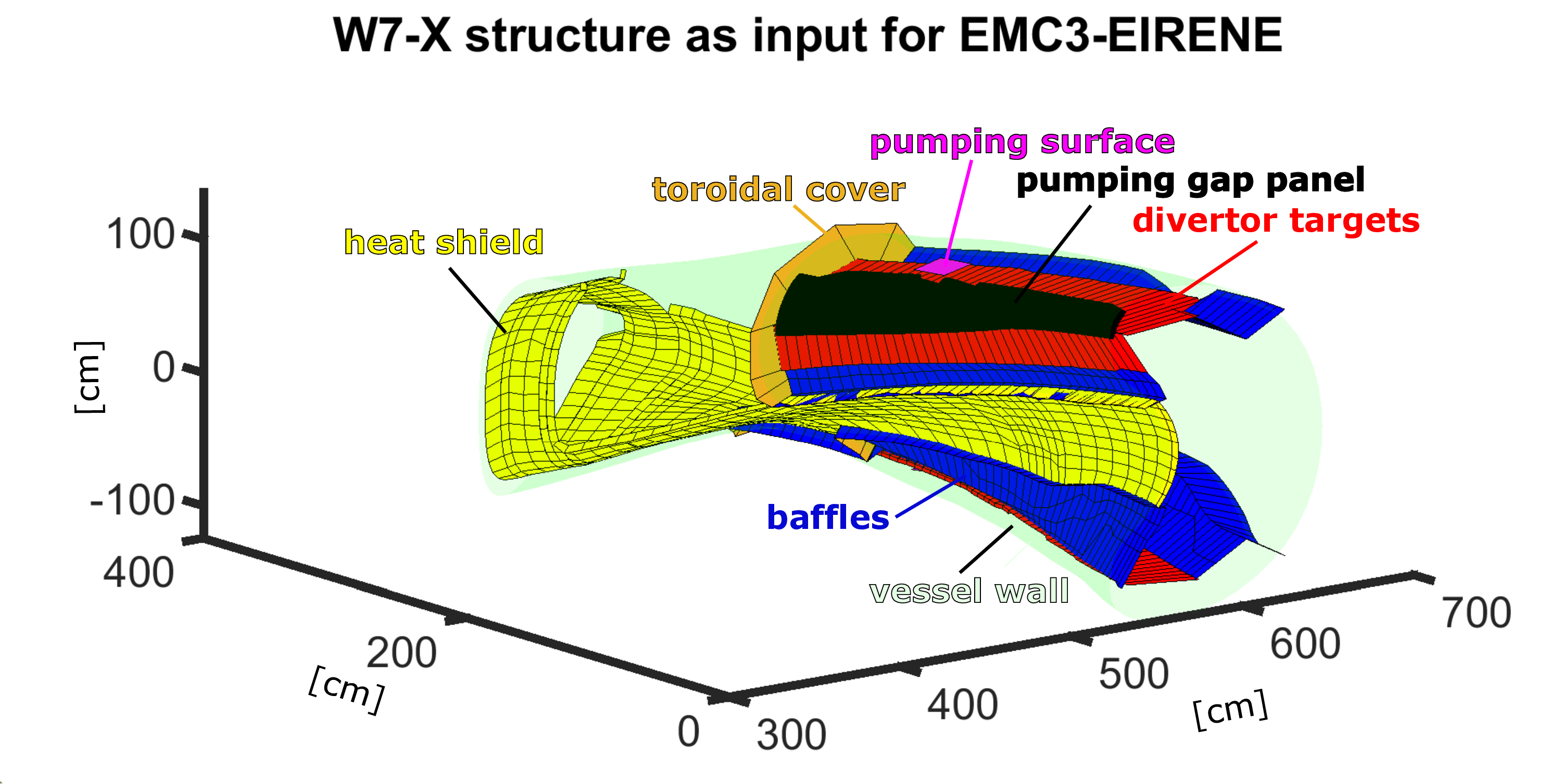}
	\caption{The included W7-X structures in the EMC3-EIRENE modeling looking from inside. The axis are expressed in centimeters.}
	\label{fig:W7Xinput}
\end{figure}

W7-X exhibits 5 fold symmetry, and each half module is a mirror image of the other half. Therefore, only half a module is simulated with EMC3-EIRENE ranging from $0^{\circ}$ to $36^{\circ}$. The structures included in the simulation are shown in figure \ref{fig:W7Xinput} including the divertor targets, baffles, heat shield, toroidal covers, vessel wall, pumping gap panel, and pumping surfaces. During OP1.2b each half module of W7-X was equipped with three turbo pumps which were connected to the subdivertor region at two locations: below the low-iota pumping gap and below the high-iota gap as indicated in figure \ref{fig:divertor_region}. In the employed EMC3-EIRENE structure, the pumps are modeled as absorbing surfaces below the actual pumping duct which is indicated in pink in figure \ref{fig:W7Xinput}. To mimic a pumping speed of $0.21\frac{\mathrm{m^3}}{\mathrm{s}}$ at the entrance of the pump ducts of W7-X, an absorption coefficient of 0.06 for the pink pumping surface is employed in the EMC3-EIRENE calculations as described in ref. \cite{feng2018evaluation}. As pumps are included in the simulation setup, but fueling not, the released neutral flux is rescaled to account for the pumped fraction. This makes that the released neutral flux is slightly larger than the ion flux reaching the targets, baffles and heath shield.

\begin{figure}[h!]
	\centering
	\begin{subfigure}{6cm}
		\includegraphics[width=6cm]{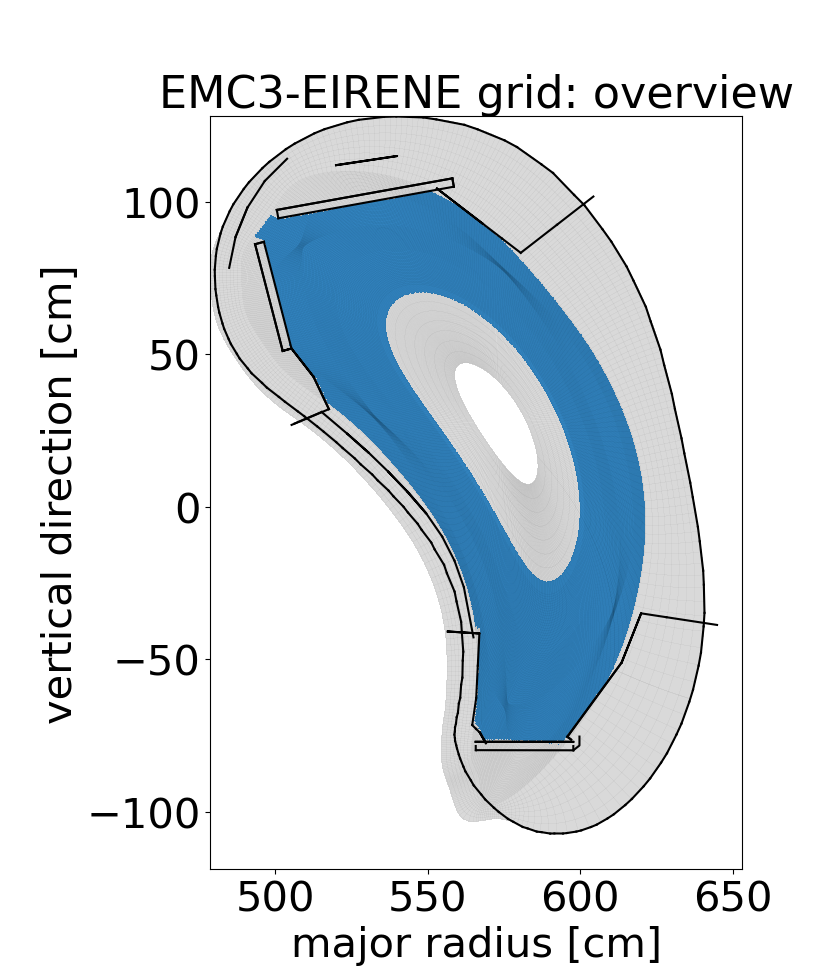}
		\caption{}
		\label{subfig:emc3grid_overview}
	\end{subfigure}
	\begin{subfigure}{6cm}
		\includegraphics[width=6cm]{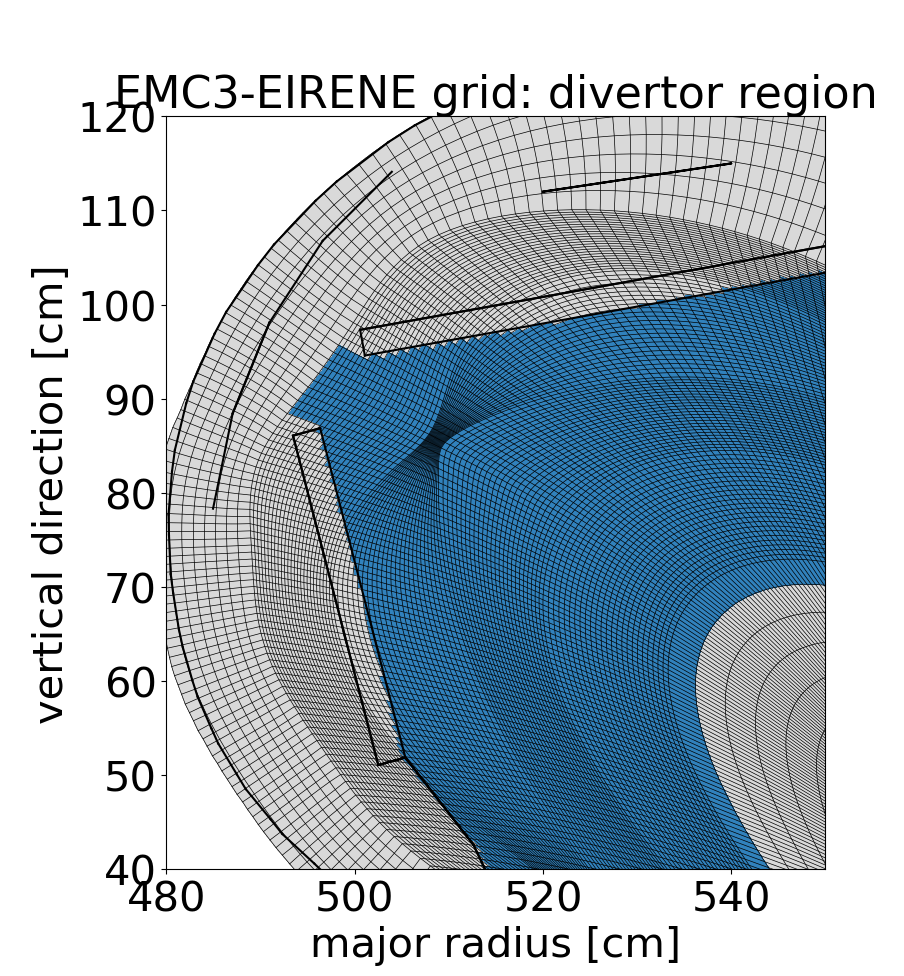}
		\caption{}
		\label{subfig:emc3grid_divertor}
	\end{subfigure}
	\caption{A toroidal cross section of the employed EMC3-EIRENE grid at $\phi = 12 ^{\circ}$. In (a) the entire cross section is shown where (b) gives a detailed view of the divertor region. The black lines indicate the in-vessel structures of W7-X which are included in the performed modeling and are also shown in figure \ref{fig:W7Xinput}. The grid part which treats both plasma and neutrals is indicated in blue.}
	\label{fig:grid_EMC3-EIRENE}  
\end{figure}

In a next step, FLARE \cite{frerichs2015field} is employed to generate a 116 x 401 x 34 (radial x poloidal x toroidal points) grid from which a toroidal cross section is shown in figure \ref{fig:grid_EMC3-EIRENE}. In the blue area the grid is both used for plasma and neutral simulations, where the remaining part only treats neutrals. The required magnetic field for grid construction is based on the vacuum field of the coils calculated by solving the Biot-Savart equations. By using the vacuum approximation for the magnetic equilibrium, the effect of plasma currents and high $\beta$ effects are not included although they influence the plasma as shown in ref. \cite{lore2019measurement}. In ref. \cite{xu2023modeling} it has been shown that these high $\beta$ effects are moderate for limited heating powers which are used in the presented work. Besides, the effects of drift flows which cause an asymmetry in the plasma between the upper and lower parts of the modules \cite{hammond2019drift, kriete2023effects} is not implemented in the current version of EMC3-EIRENE. A last effect which is not considered, are the error fields on the magnetic equilibrium. In ref. \cite{lazerson2018error} it has been shown that these effects are small, but ref. \cite{effenberg2019investigation} indicates that the effect on the plasma edge can not be ignored. 
 

As indicated in ref. \cite{feng2021understanding} the most accurate way of recreating the effect of carbon radiation in an EMC3-EIRENE simulation for detached W7-X studies, is to impose $f_{rad}$. As the discharge took place after boronization, most intrinsic impurities are bound to the wall and carbon is supposed to be the only radiating species \cite{sereda2020impact}. The imposed radiated fraction will determine the amount of carbon present in the device. For all performed simulations, $f_{rad}$ is taken from experimental bolometer data \cite{zhang2021bolometer}. As in ref. \cite{feng2021understanding}, the imposed power at the innermost boundary of the plasma grid is $90 \, \%$ of the injected power during the experiment to account for core radiation losses from Bremsstrahlung. To obtain the imposed $f_{rad}$, EMC3-EIRENE will adjust the carbon influx during the performed simulations. Going from an attached to a detached plasma state in the 3 MW simulation, this carbon influx will start at 0.101 kA in the most attached simulation for a line-integrated density of $n_{e,l.i.} = 7.0 \cdot 10^{19} \, \mathrm{m^{-2}}$, will increase to 0.156 kA where it has a rollover (for $n_{e,l.i.} = 8.5 \cdot 10^{19} \, \mathrm{m^{-2}}$) and decreases afterwards to 0.771 kA (for $n_{e,l.i.} = 10 \cdot 10^{19} \, \mathrm{m^{-2}}$). $f_{rad}$ keeps increasing though.

The electron density is imposed at the separatrix position. It is calculated from  the experimental line-integrated density obtained with the available dispersion interferometer \cite{brunner2018real}. Using the path length ($\sim 1.3 \, m$), the line-averaged density is known. In the performed study the same assumptions as in tokamaks concerning the separatrix density are made \cite{wischmeier2007simulating}. Therefore, the density at the separatrix is assumed to be $\frac{1}{3}$ of the line-averaged density. From ref. \cite{bold2022parametrisation} it can be concluded that this assumption is reasonable for W7-X. In future work, a comparison with densities measured by the Thomson Scattering system from W7-X is foreseen \cite{pasch2016thomson}. A spatially constant anomalous diffusivity of $D = 1 \, \frac{\mathrm{m^2}}{\mathrm{s}}$ is imposed both for the present hydrogen and carbon species. Besides, a spatially constant anomalous electron and ion thermal conductivity of $\chi_e = \chi_i = 3 \, \frac{\mathrm{m^2}}{\mathrm{s}}$ is assumed. All anomalous transport coefficients are taken from ref. \cite{winters2021emc3}. Ref. \cite{bold2022parametrisation} indicates that it is difficult to obtain quantitative agreement between simulations and experiments for all measured plasma and neutral quantities. However, ref. \cite{feng2021understanding} indicates that the global trends can be captured well by EMC3-EIRENE even when full agreement with experimental data is not achieved. Therefore, as in ref. \cite{winters2021emc3}, the goal is to ensure these trends rather than to obtain full agreement. In the next sections, these trends are verified by the change in the subdivertor neutral pressure, and by the $H_{\alpha}$ signal evolution.



In EMC3-EIRENE, neutral particles are traced in EMC3, while EIRENE is responsible for atomic reactions and plasma-surface interactions. 
Neutral fluxes are computed after a converged EMC3-EIRENE simulation in a post-processing simulation step. 
After a simulation is converged, one additional EIRENE iteration is made, in which fixed surfaces are added in the desired locations. The neutral fluxes to these surfaces are written out without modifying further the plasma quantities. In this way the neutral fluxes indicated in figure \ref{fig:divertor_region} are calculated in section \ref{sec:exhaust} to spatially resolve the neutrals leaving the divertor.

\section{Comparison with experimental neutral pressures}
\label{sec:comparison_exp_sim}

\begin{figure}[h!]
	\centering
	\begin{subfigure}{9cm}
		\includegraphics[width=9cm]{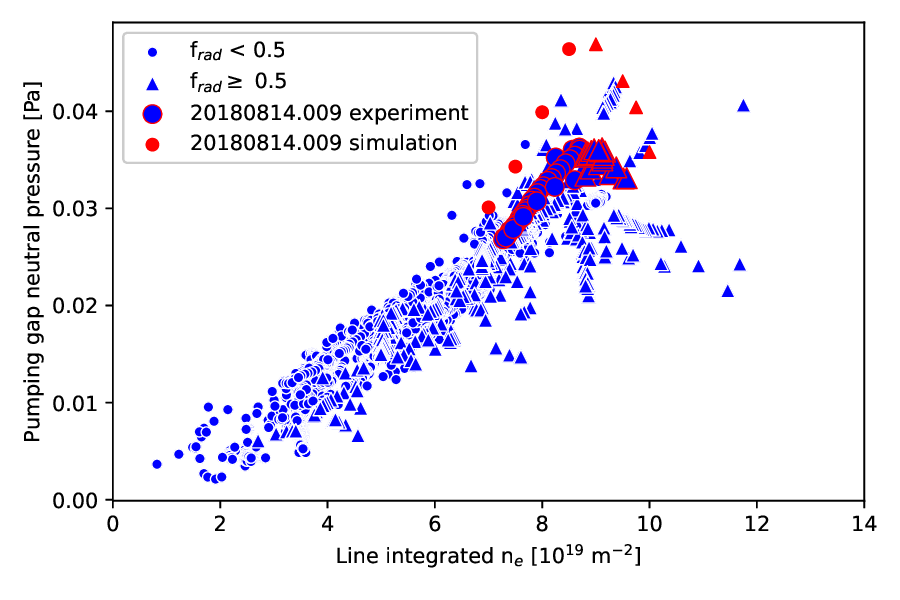}
		\caption{}
		\label{subfig:emc3eirene_pressures_3MW}
	\end{subfigure}
    \begin{subfigure}{9cm}
		\includegraphics[width=9cm]{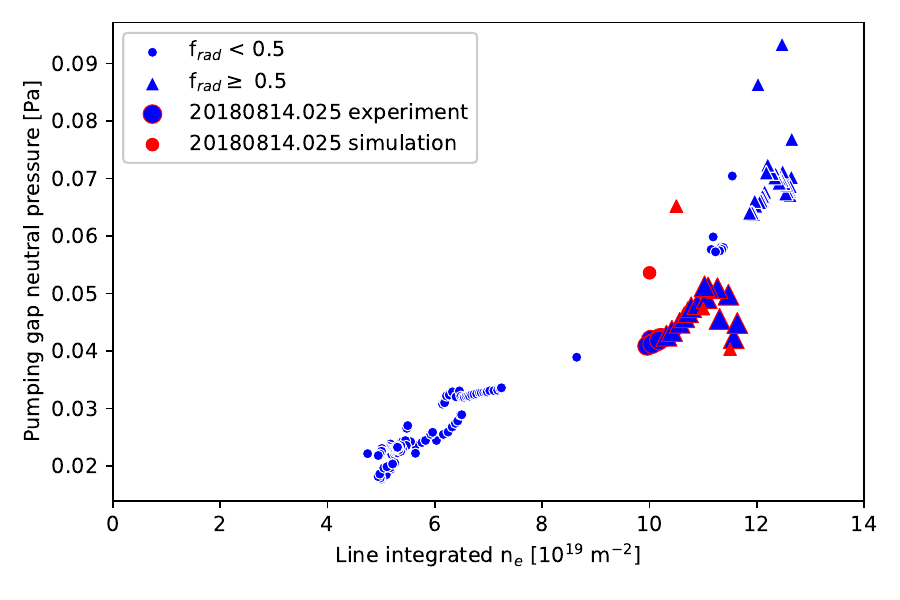}
		\caption{}
		\label{subfig:emc3eirene_pressures_5MW}
	\end{subfigure}
	\caption{Comparison between the simulated (red) and experimentally observed (blue) pressures. The simulation is tuned to the experiment shown by blue marks with red border. Data points with a radiated fraction above 0.5 are indicated with triangles, data points with a lower radiated fraction are indicated with circles. In (a) the agreement for shot 20180814.009 with an input power of 3.3 MW is shown, in (b) the agreement for shot 20180814.025 with an input power of 5.5 MW.}
	\label{fig:emc3eirene_pressures}  
\end{figure}

The investigated experiments are programs 20180814.009 with an injected electron cyclotron resonance heating of $\sim$ 3.3 MW and 20180814.025 with $\sim$ 5.5 MW \cite{schmitz2020stable}. The full set of available diagnostics to which the simulations can be compared for programs in OP1.2b, can be found in ref. \cite{hathiramani2018upgrades}. In a first step, the modeled subdivertor pressures are compared with the experimental ones at the vessel wall below the pumping gap \cite{wenzel2019performance}. 
A comparison with this measurement is shown in figure \ref{fig:emc3eirene_pressures} for both simulations. The simulated pressures are not only compared with the experimental data of ref. \cite{schmitz2020stable} but with all similar experiments during OP1.2b in standard magnetic configuration (after boronization). In program 20180814.009 a larger density range is achieved than in  experiment 20180814.025. As the radiated fraction is taken from experiments, also the simulation frame is limited to the densities from the experiment. This makes that it was possible to perform more simulations for the first case making it easier to verify trends between simulations and experiments. On top, it was more difficult to reach steady-state required for the bolometer data for the investigated densities in the 5.5 MW experiment than in the 3.3 MW program limiting the number of 5.5 MW simulations to four.


In this comparison, the measured neutral pressure is compared with the simulated molecular pressure. In fact, the atoms are thermalized in the ducts of the pressure gauge. This decreases the atom temperature. As the pressure measurement relies on the temperature of neutrals \cite{haas1998in}, the contribution of the atomic pressure to the measured pressure can be ignored. 

\begin{figure}[h!]
	\centering
	\begin{subfigure}{6cm}
		\includegraphics[width=6cm]{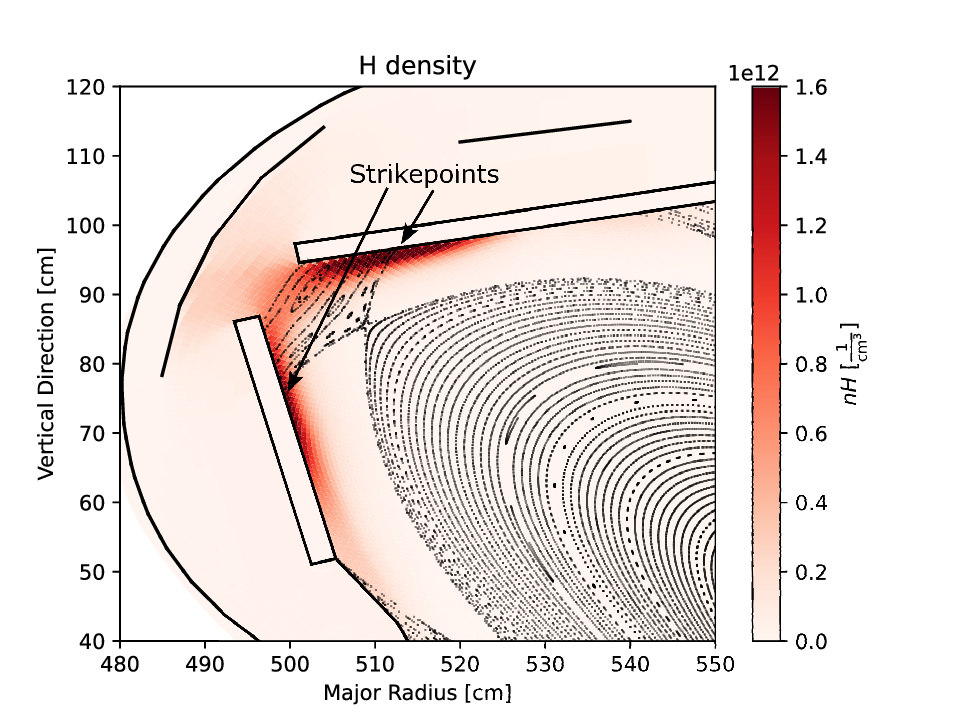}
		\caption{}
		\label{subfig:atom_density_attached}
	\end{subfigure}
	\begin{subfigure}{6cm}
		\includegraphics[width=6cm]{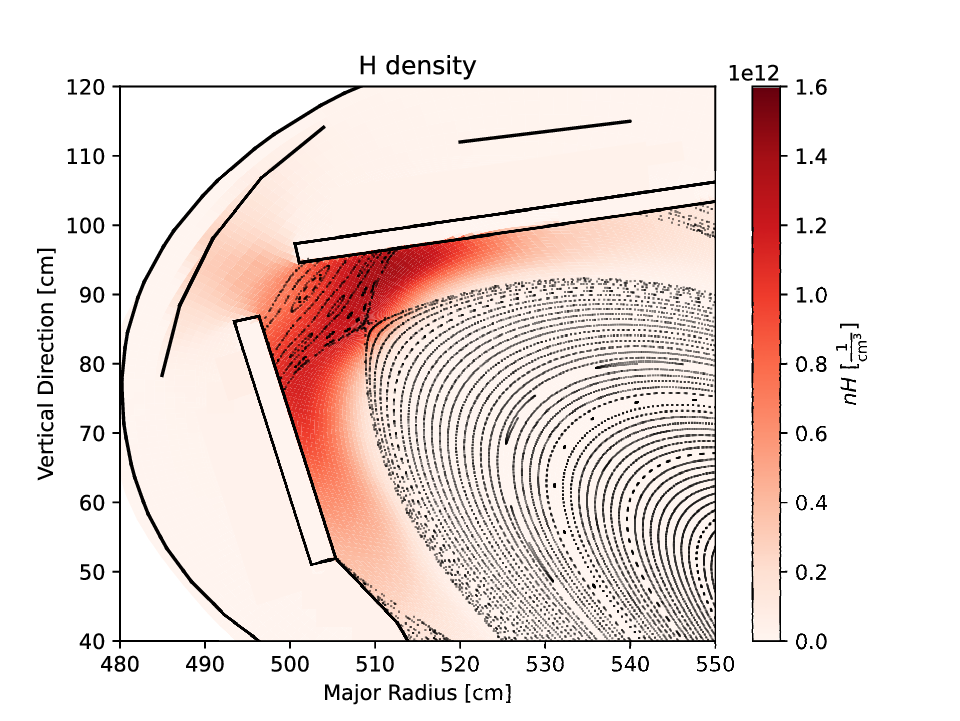}
		\caption{}
		\label{subfig:atom_density_detached}
	\end{subfigure}
	\caption{The simulated hydrogen atom densities for attached (a) and detached (b) conditions in the vicinity of the divertor target. In attached conditions the density is higher close to the targets but decreases fast where in detached conditions this is a smoother transition. On the right plot, the Poincaré curves are added to highlight the structure of the island divertor in W7-X.}
	\label{fig:atom_density}  
\end{figure}

Figure \ref{fig:atom_density} shows the simulated neutral hydrogen atom density profiles in the vicinity of the divertor with a line-integrated density of $7 \cdot 10^{19} \mathrm{m}^{-2}$ on the left (attached, lowest simulated density in figure \ref{subfig:emc3eirene_pressures_3MW}), and $10 \cdot 10^{19} \mathrm{m}^{-2}$ on the right (detached, highest simulated density in figure \ref{subfig:emc3eirene_pressures_3MW}) for the case with 3 MW of input power. Under attached conditions, the sharp density gradient towards the plasma core on figure \ref{subfig:atom_density_attached} shows that a lot of the generated neutrals are ionized in the vicinity of the divertor targets and that the ionization front is moved more inwards with detachment. Ref. \cite{feng2021understanding} shows that the neutral penetration length in the near-target region increases when the amount of radiation is increased due to this inward shift. On top, the influence of the charge exchange processes and the elastic collisions will modify the path neutrals travel in the reactor leading to longer mean-free paths which makes that more of them reaches the pumping gap as shown in the next sections \cite{feng2021understanding}. On figure \ref{fig:atom_density}, the Poincar{\'e} plot illustrates that the location of high ion fluxes - and in that way the one of high neutral density - falls together with the strikepoint location. The decrease in hydrogen atom density behind the pumping gap is explained by 
an increase of hydrogen molecules in this region. Due to the interaction from the atoms with the wall and subdivertor structures in the stellarator, they form hydrogen molecules.

\section{Location of neutralization}
\label{sec:origin_neutrals}

As mentioned in the introduction, neutrals in the W7-X experiments 
originate from interaction between plasma and plasma facing components (mainly divertor targets and baffles). 

For the simulation with 3 MW of input power, in attached conditions, $81.2 \, \%$ of the neutrals are born through plasma wall interaction at the divertor targets -- $54.1 \, \%$ due to interaction on the horizontal divertor and $27.1 \, \%$ due to interaction on the vertical one --, $17.5 \%$ due to interaction at the baffles, and $1.3 \, \%$ due to interaction at the heat shield. This agrees with the findings of ref. \cite{winters2021emc3} where the small influence of the heat shield in the neutralization process is highlighted. In detached conditions, the neutrals are spread over a larger region. Only $75.8 \, \%$ of them are born at the targets, $22.7 \, \%$ at the baffles, and $1.6 \, \%$ at the heat shield. In this detached simulation, less neutrals are generated: the number of them born through plasma wall interaction decreases with $57.4 \, \%$.

\begin{figure}[h!]
	\centering
	\begin{subfigure}{8cm}
		\includegraphics[width=8cm]{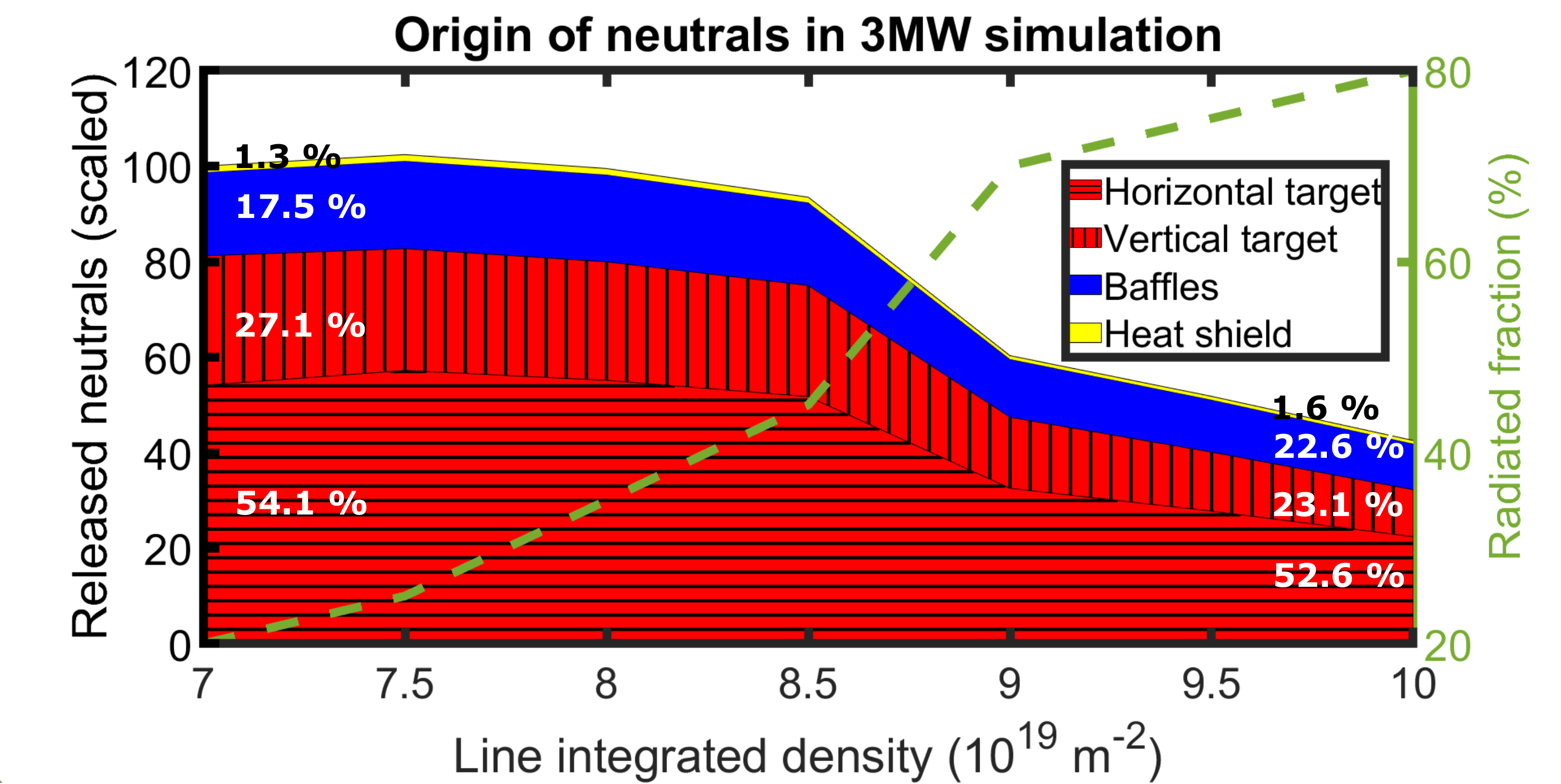}
		\caption{}
		\label{subfig:origin_neutrals_3MW}
	\end{subfigure}
	\begin{subfigure}{8cm}
		\includegraphics[width=8cm]{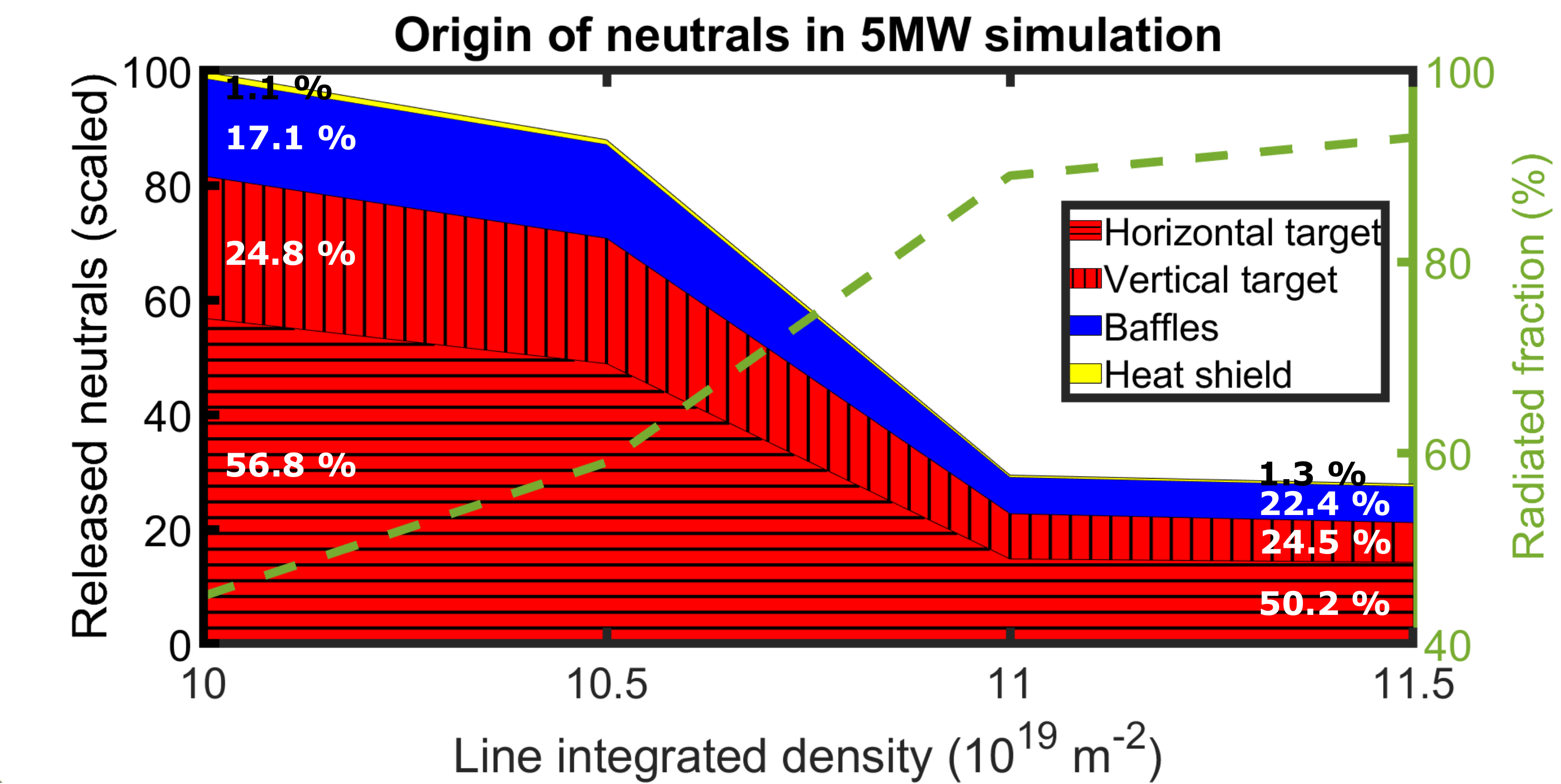}
		\caption{}
		\label{subfig:origin_neutrals_5MW}
	\end{subfigure}
	\caption{Overview of the origin of the neutrals in the performed simulations for the 3 MW (a) and 5 MW (b) simulation.}
	\label{fig:origin_neutrals}  
\end{figure}

The simulation with 5 MW of input power shows the same overall trends. Both simulation sets make clear that the generated neutrals are decreasing significantly when the radiated power fraction (and the line-integrated density) increases which is in line with the findings of refs. \cite{feng2021understanding, jakubowski2021overview, perseo20212d}. In ref. \cite{jakubowski2021overview} the same 5 MW discharge is analyzed. Over there, the same trends are observed by the divertor Langmuir probes: the radiation increases from $40 \, \%$ to $90 \, \%$ where the particle flux (and in that way the number of generated neutrals) decreases from $\sim 2.3 \frac{\mathrm{A}}{\mathrm{m^2}}$ to $\sim 0.9 \frac{\mathrm{A}}{\mathrm{m^2}}$. An overview from all the generated neutrals in the performed simulations is given in figure \ref{fig:origin_neutrals}.

The simulations are only compared at few locations with experiments. On top, the sensitivity of the performed simulations towards the not exactly known input parameters (like radiated fraction) is not investigated, and the numerical errors are not analyzed in detail. Therefore, no error bars on the simulation result are defined and the percentages coming from the simulations should be taken qualitatively rather than quantitatively.

\begin{figure}[h]
	\centering
	\includegraphics[width=12cm]{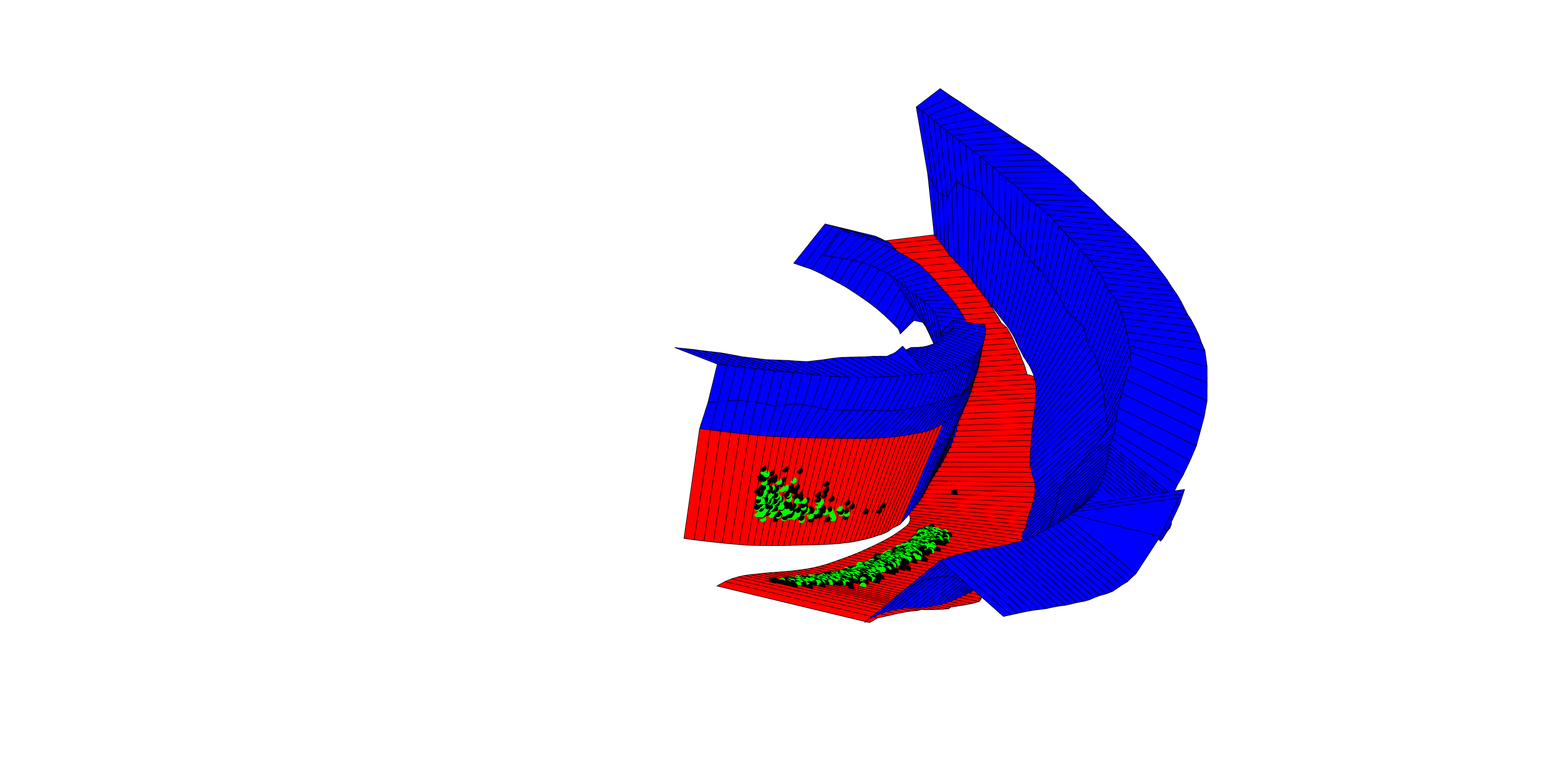}
	\caption{The strikepoints under attached (green) and detached(black) conditions. Especially on the vertical target it is clear in the figure that the black points (detached) are spread over a larger area in comparison with the green ones (attached).}
	\label{fig:strikepoints}
\end{figure}

Figure \ref{fig:origin_neutrals} indicates that the distribution of neutral sources is only slightly influenced by the regime of the plasma, both in attached and detached simulations the main  neutral source are the targets. Under detached conditions, however, there is a slightly increase of generated neutral fraction at the baffles. This can be explained by the strikepoint plots of figure \ref{fig:strikepoints}. For these strikeline plots, FLARE is employed. The diffusion coefficient in FLARE takes into account the temperature of the particles at the divertor target resulting in a larger diffusion for colder cases. This makes that under detached conditions in which the temperature at the divertor target is colder, the strikepoints are more spread out and coming closer to the baffles in comparison with field line following under attached conditions. Therefore, a larger fraction of the recycling flux will originate from the baffles. By comparing figure \ref{subfig:origin_neutrals_3MW} and \ref{subfig:origin_neutrals_5MW} no influence of the input power on the recycling flux location is noticed.

\begin{figure}[h]
	\centering
	\includegraphics[width=8cm]{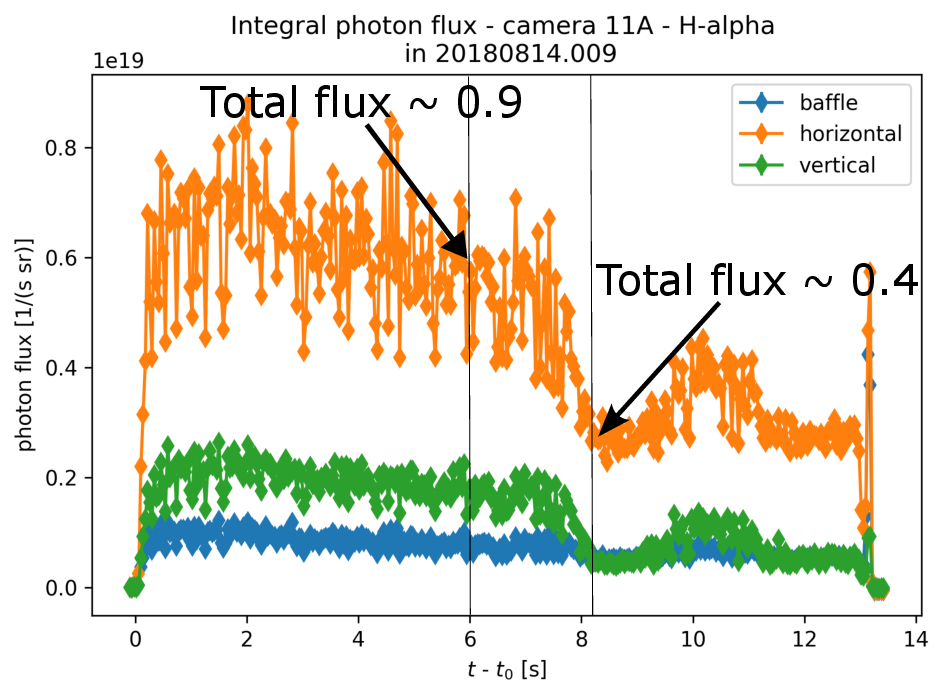}
	\caption{The $H_{\alpha}$ intensity for discharge 20180814.009 which is the modeled shot with 3 MW of input power.}
	\label{fig:H_alpha}
\end{figure}

For attached conditions, ref. \cite{kremeyer2022analysis} discusses the origin of neutrals in W7-X discharges based on experimental observations. As in the presented simulations, the majority of the recycled neutrals originates from the horizontal divertor. For the examined 3 MW shot (experiment 20180814.009) one of the $H_\alpha$ signals is given in figure \ref{fig:H_alpha}. For the other $H_\alpha$ cameras similar signals were obtained (both in the $3MW$ and $5 MW$ programs). The $H_\alpha$ signal at 6 s agrees with a line-integrated density of $\sim 7.0 \cdot 10^{19} \mathrm{m^{-2}}$, and at 8.2 s with a line-integrated density of $\sim 9.5 \cdot 10^{19} \mathrm{m^{-2}}$. As indicated on the figure, the flux decreases with $\sim 56 \, \%$ during this transition. This agrees well with the simulation in which the flux decreases with $48.1 \, \%$ between the two simulations as can be seen in figure \ref{fig:origin_neutrals}. 
As in the simulations the majority of neutrals originates from the horizontal divertor, especially from the vicinity of the strikepoint. In the performed simulations the spatial distribution of the generated neutrals was studied for the simulations with 3 MW of input power and also shows that most of them originate from the strikeline around the low-iota pumping gap. 

\section{Exhaust of the neutrals out of the divertor region}
\label{sec:exhaust}

 In a fusion device the goal is to pump as little main neutrals as possible, but enough to keep density control and as much as necessary to remove helium. Otherwise, the temperature throughput becomes a problem. In this section, it is analyzed what is happening with the generated hydrogen neutrals of section \ref{sec:origin_neutrals}.
 
\begin{figure}[h!]
	\centering
	\begin{subfigure}{6cm}
		\includegraphics[width=6cm]{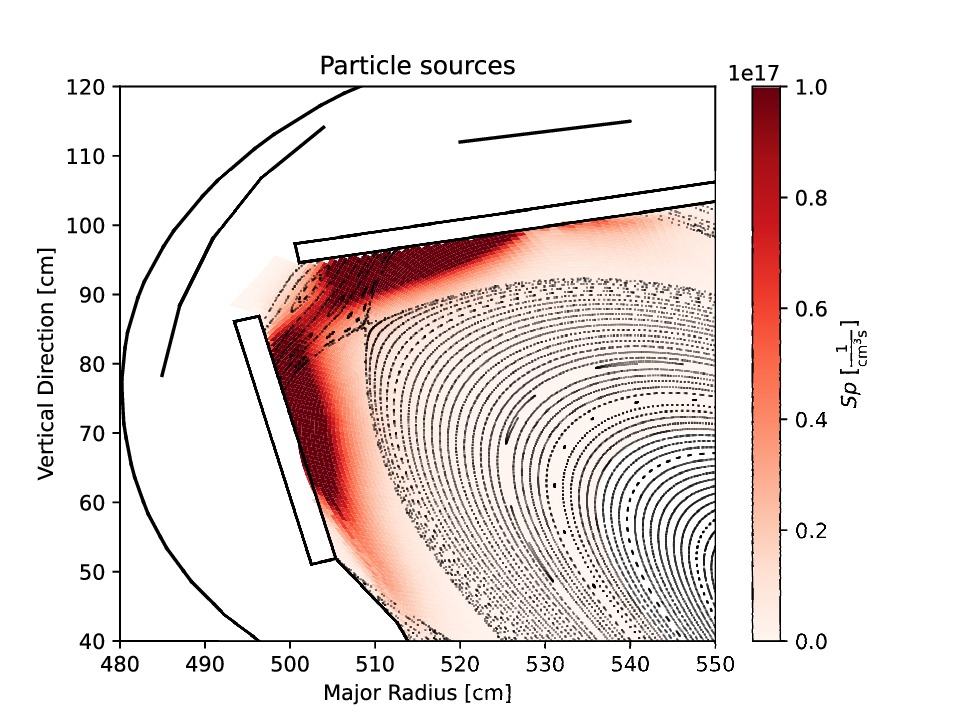}
		\caption{}
		\label{subfig:sources_attached}
	\end{subfigure}
	\begin{subfigure}{6cm}
		\includegraphics[width=6cm]{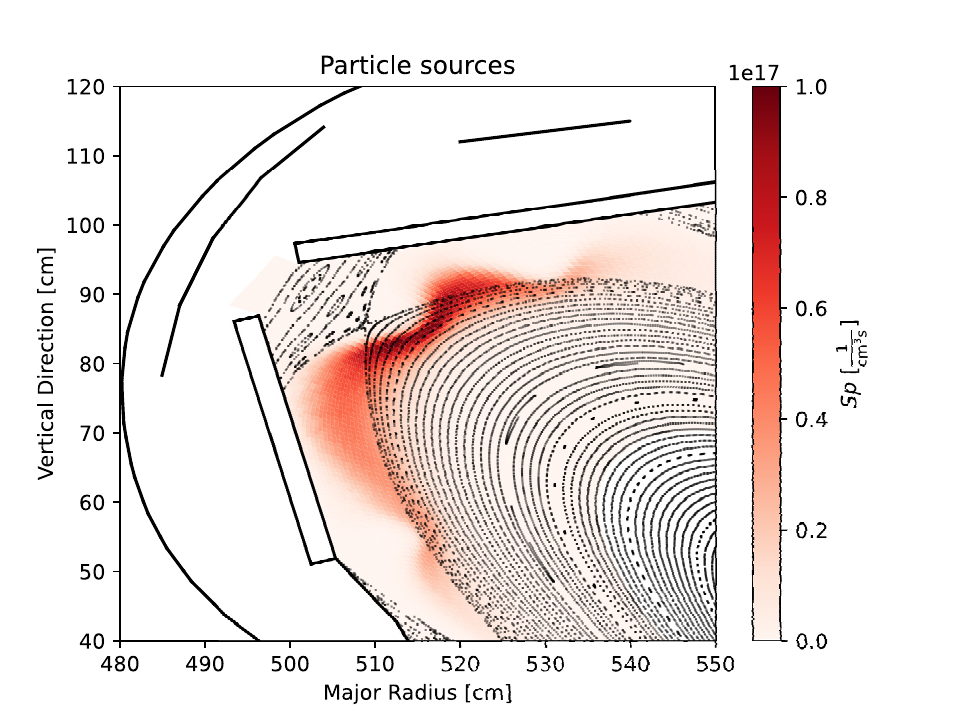}
		\caption{}
		\label{subfig:sources_detached}
	\end{subfigure}
	\caption{The simulated ionization sources for the lowest density, attached simulation (a), and for the highest density, detached simulation (b) in case of 3 MW input power.}
	\label{fig:sources}  
\end{figure}

After being born through plasma wall interaction, the majority of the neutrals will be re-ionized 
in the divertor region. Under attached conditions more particles are ionized in comparison with under detached conditions as can be seen in figure \ref{fig:sources}. This is partially caused because more particles are generated in attachment (see figure \ref{fig:origin_neutrals}), but also because in detachment the neutrals will travel further as demonstrated in ref. \cite{feng2021understanding}. This causes more particles to leave the divertor region during detached plasma operation, but also explains why the ionization front moves inwards in detachment.

Besides re-ionizing, a part of the particles will be collected at the pumping gap, and another fraction of the neutrals will also leave the divertor region in the poloidal and toroidal direction into the main chamber, as indicated in figure \ref{fig:divertor_region}. In order to maximize the subdivertor pressure, and in that way the pumping efficiency, it is important to keep these poloidal and toroidal fluxes as low as possible, the divertor plugging should be as high as possible. 

\begin{figure}[h!]
	\centering
	\begin{subfigure}{8cm}
		\includegraphics[width=8cm]{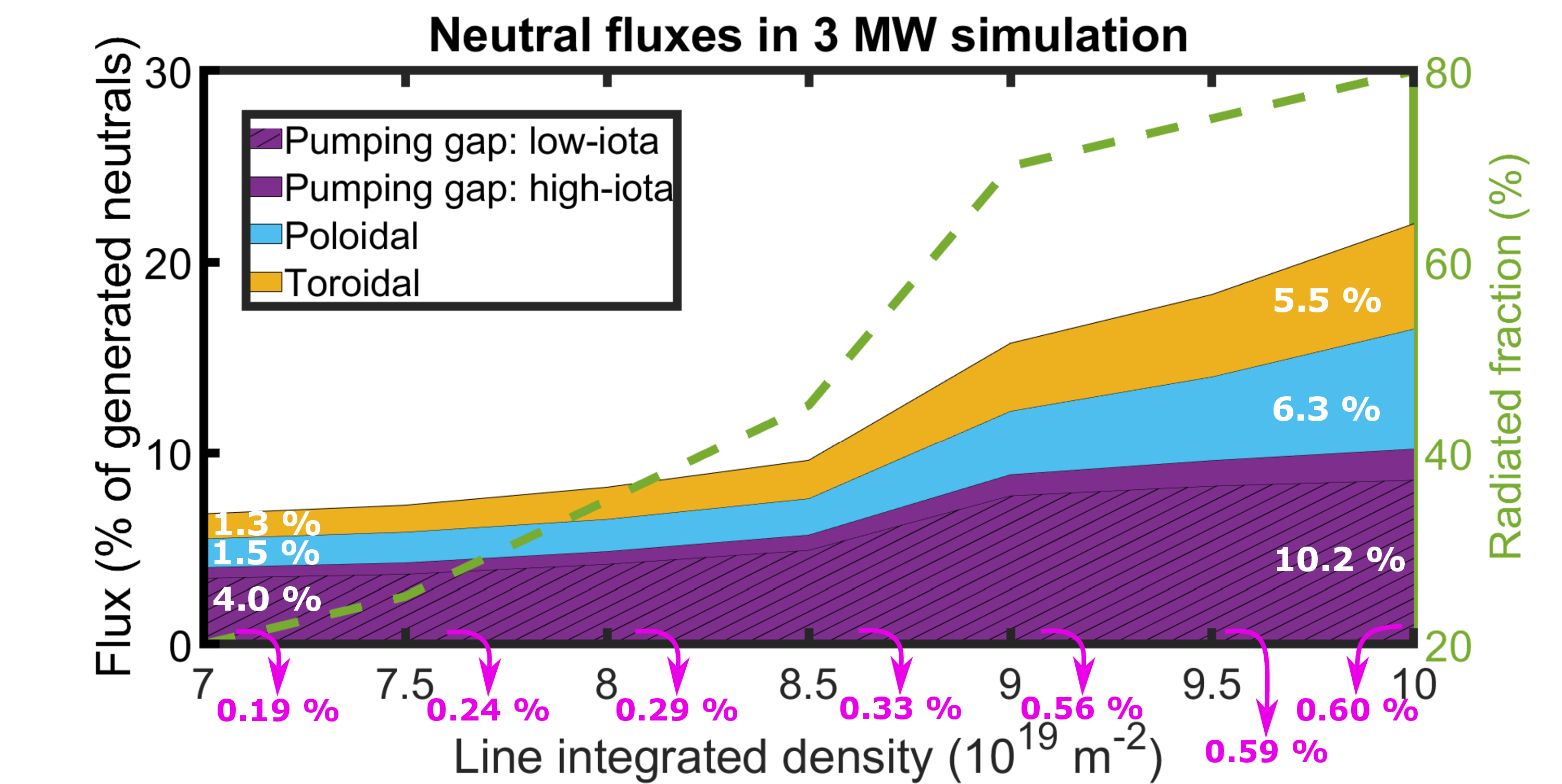}
		\caption{}
		\label{subfig:neutral_flux_3MW}
	\end{subfigure}
	\begin{subfigure}{8cm}
		\includegraphics[width=8cm]{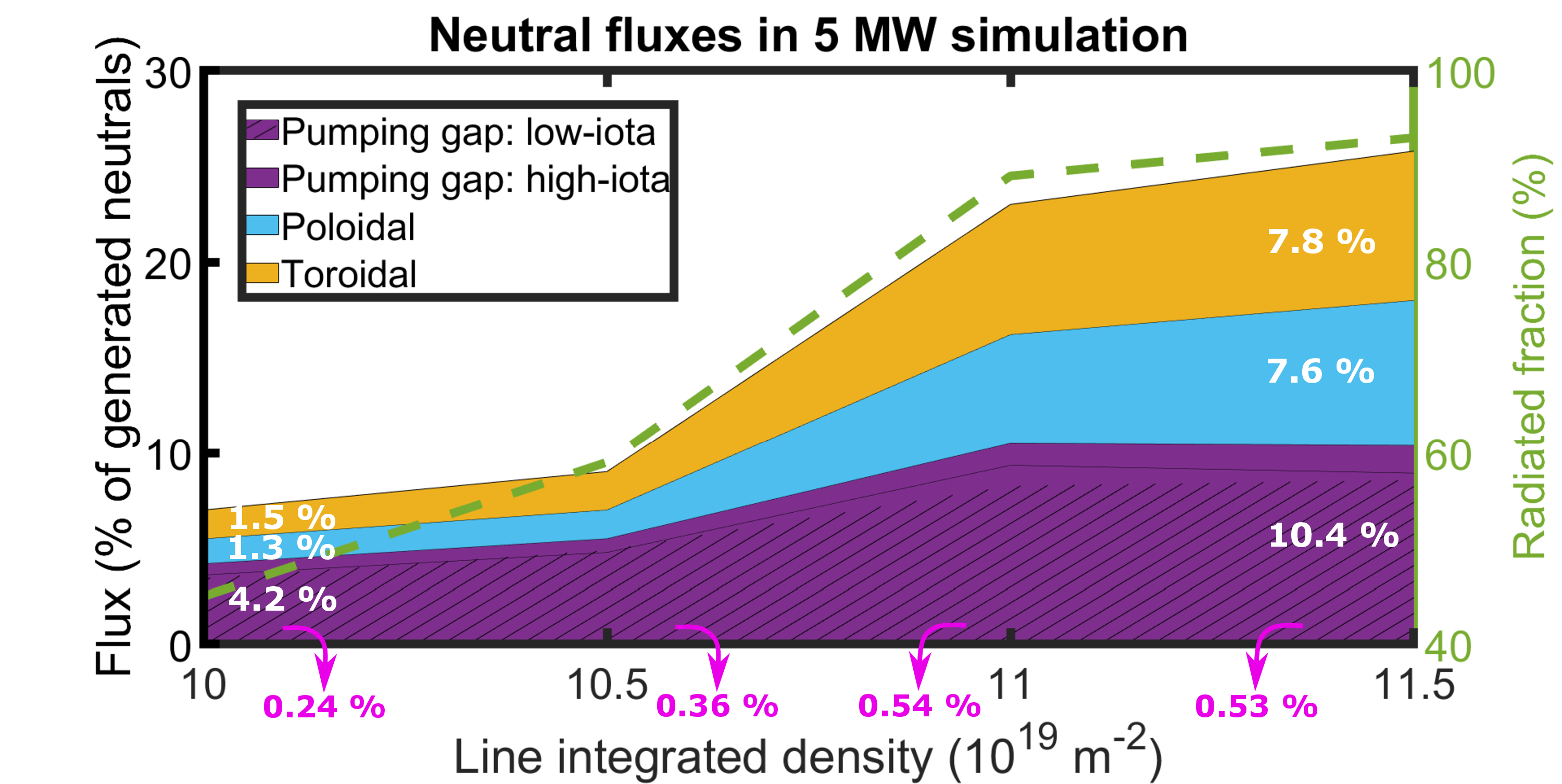}
		\caption{}
		\label{subfig:neutral_flux_5MW}
	\end{subfigure}
	\caption{Overview of the neutral fluxes in the performed simulations for the 3 MW (a) and 5 MW (b) simulation. The pumped fraction is indicated in pink as a percentage of the total number of neutrals. The particles which make it through the pumping gap, but are not pumped, flow back through the pumping gap to the main chamber. It should be remarked that the number of released neutrals is decreasing while increasing the radiative fraction as indicated in figure \ref{fig:origin_neutrals}.}
	\label{fig:neutral_flux}  
\end{figure}

The EMC3-EIRENE analysis indicates that for 3 MW of input power, in attached conditions, $1.5\%$ of the neutrals neutralized at the divertor is leaving the divertor region in the poloidal direction, $1.3\%$ in the toroidal direction, $4.0\%$ of the particles is collected at the pumping gap, and the remaining part is re-ionized in the divertor region. In this analysis, the neutrals are only followed from the divertor targets until they reach the border of the divertor region (see figure \ref{fig:divertor_region}). An extended analysis in which the neutrals are also followed afterwards which makes it possible to determine how many neutrals are coming back to the divertor region is left for future work. Similar flux ratios are observed in attached conditions with 5 MW input power as can be seen in figure \ref{fig:neutral_flux}. 


The performed analysis shows that only limited number of neutrals is guided to the subdivertor which is to be expected for an open divertor. 
On top, a large part of the neutrals reaching the pumping gap will eventually leak back into the plasma. For the performed simulations, the pumped flux is analyzed. In the 3 MW attached simulation, only $0.19 \, \%$ of the generated neutrals is pumped out. In the EMC3-EIRENE model, this means that only $0.19 \, \%$ of the particles is absorbed by the pink surface of figure \ref{fig:W7Xinput}. Given that $4.0 \%$ of the particles is reaching the pumping gap, this means that $95 \, \%$ of the particles entering the pumping gap are leaking back into the plasma. In the performed simulations the only gap through which the neutrals leak back into the plasma is the pumping gap. In reality, also small gaps exist between the divertor tiles and the leaking fraction will be higher. 

These findings agree with the experimental estimates from ref. \cite{wenzel2022gas}. Over there the net flow through the pumping gap ($\Gamma_{gap}$) is investigated. This is the flux reaching the pumping gap minus the neutral flux coming from the subdivertor and going back into the plasma. The studied discharges differ between ref. \cite{wenzel2022gas} and the work presented in this paper. Nevertheless, the order of magnitude for the pumped flux agrees, but $\Gamma_{gap}$ is an order of magnitude smaller than the flux towards the divertor shown in figure \ref{fig:neutral_flux} from which the backflow is not subtracted. In OP1.2b only turbo pumps were available. As indicated in ref. \cite{schlisio2021evolution} it is expected that newly installed cryo pumps increase the pumped fraction, even when the subdivertor pressure stays similar.

The pumping gap can be split in two regions: the low-iota pumping gap, and the high-iota pumping gap which is located further down the divertor far away from the strikeline location in standard configuration. As expected the low-iota pumping gap collects most of the pumped flux. The performed EMC3-EIRENE simulations show, that this is independent of density, injected power, and radiation. In all examined cases around $84 \, \%$ of the flux to the subdivertor region is going through the pumping  gap at the low-iota end of the target.

The number of particles escaping in the poloidal and toroidal direction, is an indication how well the particles are plugged to the divertor. As only $2.8\%$ of the particles leaves the divertor region in the poloidal and toroidal direction, the neutrals are plugged well to the divertor in attached conditions. 


In detachment, plugging is decreasing and $11.8\%$ of the particles leaves the divertor in the poloidal or toroidal direction for the 3 MW simulation, and even $15.4 \, \%$ for the 5 MW simulation. The inwards move of the ionization front explains why the plugging decreases in detachment. The main problem, however, is the limited number of neutrals gathered by the turbo pumps. During detachment the pumped flux slightly increases, but due to the low pressures, the flow stays in the molecular flow regime as the subdivertor pressure stays too low to enhance neutral neutral collisions. This means that particles bounce from wall to wall and are not guided to the pumps by friction from neutral-neutral collisions. As a result, the pumped flux stays below one percent and the total pumped fraction even drops as the number of generated neutrals is lower. An improved subdivertor structures which guides the neutrals better to the pumps and avoids that they flow back into the main chamber can improve the pumped fraction as up to $10.4 \, \%$ of the generated neutrals makes it to the pumping gap. 

Figure \ref{fig:neutral_flux} shows that a four to six times higher fraction of the neutrals escape in the poloidal or toroidal direction in detachment in comparison with attachment. How much higher the fraction is, seems to depend on the input power. Furthermore, the collection fraction in detachment doubles in comparison with the one in attachment. This seems to be independent of the input power. This also results in a double fraction of pumped out particles.

\begin{figure}[h!]
	\centering
	\begin{subfigure}{6cm}
		\includegraphics[width=6cm]{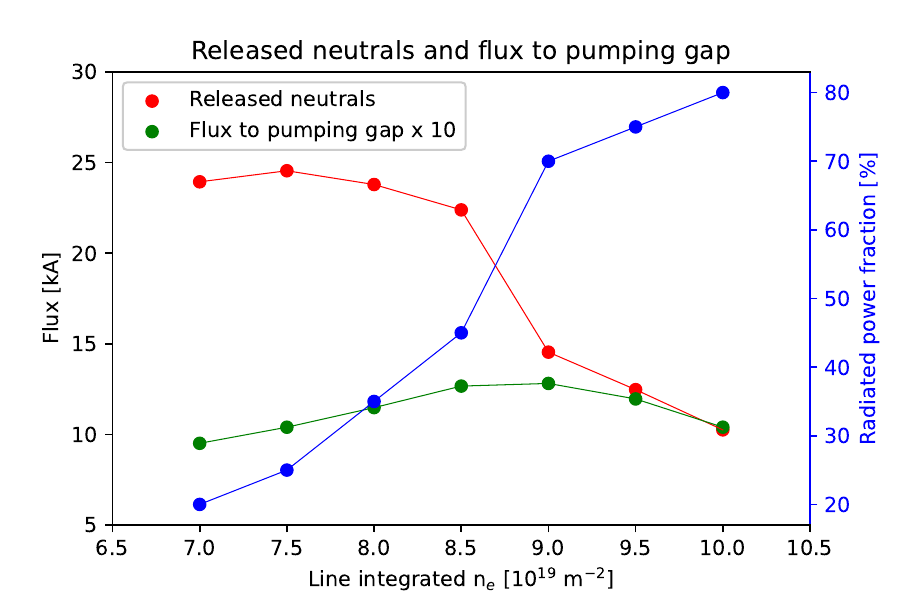}
		\caption{}
		\label{subfig:fluxes_3MW}
	\end{subfigure}
	\begin{subfigure}{6cm}
		\includegraphics[width=6cm]{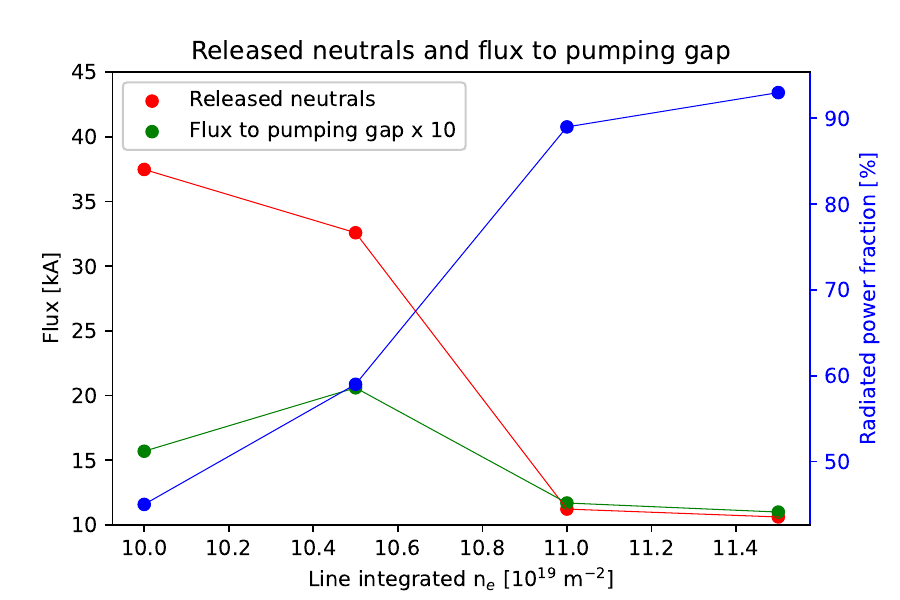}
		\caption{}
		\label{subfig:fluxes_5MW}
	\end{subfigure}
	\caption{The total released neutral flux and the fraction reaching the pumping gap for the simulations with 3 MW (a) and 5 MW (b) of input power. On the right axis the corresponding radiated fraction is indicated.}
	\label{fig:fluxes}  
\end{figure}

The performed simulations show that the number of generated neutrals increases for small line-integrated densities (and small radiative fractions) until it reaches a saturation point after which the incoming ion flux and in that way the number of released neutrals starts to decrease. Figure \ref{subfig:fluxes_3MW} shows that this maximum is achieved for a line-integrated density of $7.5 \cdot 10^{19} \mathrm{m}^{-2}$ ($f_{rad} = 0.25$) in case of 3 MW of input power. When the released neutral flux start to decrease, the transition to detachment has started \cite{feng2021understanding} after which $f_{rad}$ starts to control the behavior of the discharge rather than the imposed $n_{e,l.i.}$. For intrinsic impurities it has been demonstrated in ref. \cite{perseo20212d} that $f_{rad}$ rather scales with $n_{e}^2$. The main decrease in released neutrals takes place when $f_{rad}$ exceeds 0.5 as can be seen on the figure. Between $n_{e,l.i.} = 8.5 \cdot 10^{19} \mathrm{m}^{-2}$ and $n_{e,l.i.} = 9.0 \cdot 10^{19} \mathrm{m}^{-2}$, the radiated fraction increases from $45 \, \%$ to $70 \, \%$. The maximum number of particles are collected for a line-integrated density of $9.0 \cdot 10^{19} \mathrm{m}^{-2}$. Ref. \cite{feng2021understanding} explains the later peak in flux to the pumping gap in comparison with the peak in neutral flux due to the increased neutral penetration length while detaching the divertor. From figure \ref{fig:fluxes} (which presents the simulation results from before in a different way) the percentage of generated neutrals reaching the divertor gap -- the green curve divided by the red one -- always increases with the line-integrated density, also after the rollover for the flux to the pumping gap. Figure \ref{subfig:fluxes_5MW} shows a similar behavior for the simulations with 5 MW of input power. 

\section{Summary and conclusion}

In future fusion power plants, density control and exhaust of neutral particles are crucial for operation. To understand better the particle exhaust in W7-X, EMC3-EIRENE simulations are performed to analyze the generation of neutrals and fluxes in  the divertor region of W7-X. The performed simulations cover a range of line-integrated densities in standard configuration for different heating powers and radiated fractions. In a first step, the simulated neutral subdivertor pressures are compared with experimental observations. The same trends in neutral pressures are found between experiments and the performed simulations. This allows to study the overall trends with EMC3-EIRENE simulations.

An analysis of the origin of the neutrals show that the baffles become more important for the recycling flux while going from attached conditions to detached ones. In detachment the ion flux arriving at the divertor region is decreasing meaning that less neutrals are generated. These findings are independent of the input power. Currently, only one configuration, the so-called standard configuration, is investigated. It has been shown in ref. \cite{stephey2018impact} that the magnetic configuration can influence the exhaust of hydrogen and helium. An in-depth analysis of the neutral fluxes with EMC3-EIRENE in such other configurations is left for future work.

After studying the origin of the neutrals, the EMC3-EIRENE simulations are used to determine where the neutrals are going. In attached conditions a large fraction re-ionizes close to the divertor targets. Due to a longer neutral penetration length, the ionization front will move inwards in detachment and a smaller fraction of the neutrals re-ionizes. The flux captured at the pumping gap increases, but also a larger fraction of the particles escapes the divertor region in the poloidal and toroidal directions. 

The pumped flux increases accordingly, but as the subdivertor pressures do not increase, only $0.19 \, \%$ of the neutrals is pumped out. The increase of the fraction of the neutrals leaving in toroidal and poloidal directions shows that detachment is in the current target geometry good for heat exhaust, but has draw back when it comes to plugging the neutrals to the divertor region. As in all simulations less than $8 \, \%$ of the neutrals is leaking in the toroidal direction, the performed simulations demonstrate that a toroidally segmented divertor is feasible. The main gain in pumped fraction, however, can be achieved by guiding the particles which make it to the pumping gap in a better way to the pumps and in that way avoid that they leak back to the main chamber.


\section*{Acknowledgements}

This work was funded by the Department of Energy under grant number DE-SC0014210.

The publisher, by accepting the article for publication acknowledges, that the United States Government retains a non-exclusive, paid-up, irrevocable, world-wide license to publish or reproduce the published form of this manuscript, or allow others to do so, for United States Government purposes. 

This work has been carried out within the framework of the EUROfusion Consortium, funded by the European Union via the Euratom Research and Training Programme (Grant Agreement No 101052200 - EUROfusion). Views and opinions expressed are however those of the author(s) only and do not necessarily reflect those of the European Union or the European Commission. Neither the European Union nor the European Commission can be held responsible for them.

The EMC3-EIRENE and FLARE simulations were performed at the UW Madison Centre for High Throughput Computing (CHTC) https://doi.org/10.21231/GNT1-HW21 and on the HPC system Raven at the Max Planck Computing and Data Facility.

\bibliography{bib_emc3_w7x}

\end{document}